# Fluctuations of the aperture-averaged orbital angular momentum after propagation through turbulence

## MIKHAIL CHARNOTSKII[1,*]

[1] *Erie, CO, 80516*
*Corresponding author: Mikhail.Charnotskii@gmail.com*

**In the recent paper [1] it was shown that for paraxial propagation of scalar waves, the transverse linear momentum and orbital angular momentum (OAM) are related to the wave coherence function. Although both of these quantities are conserved during free-space propagation, they fluctuate for beam propagation in a random inhomogeneous medium. We hereby present an extension of this theory to the case of OAM fluctuations of a spherical wave intercepted by a finite aperture. A complete asymptotic theory for the aperture-averaged OAM variance is developed for both weak and strong fluctuation conditions, based on the asymptotic expansions of the Feynman path-integral solution for the fourth-order coherence function of a spherical wave propagating through a random inhomogeneous medium. We show that the OAM fluctuation averaging is very different for circular-symmetric and non-circular-symmetric apertures. Aperture averaging of OAM fluctuations does not obey the standard "square root" law, and that the weak/strong fluctuation conditions for the aperture-averaged OAM are not defined by the values of the scintillation index of the incident wave.**

## 1. INTRODUCTION

It is shown in [1] that the orbital angular momentum (OAM) of an optical wave is related to the coherence function of the propagating wave. This makes it possible to use a well-developed parabolic equations technique for the coherence functions of various orders [2] to investigate the OAM of waves propagating through atmospheric turbulence.

The mean value and variance of the total OAM for several types of beam waves were investigated in [3–6]. In [1], the mean and variance of the total beam TLM and OAM were investigated based on the Markov Approximation [2] for statistical moments of wave propagating in a random inhomogeneous medium. It was shown that while scattering by medium inhomogeneity causes TLM and OAM fluctuations, the mean values are conserved. It was also determined that the total OAM fluctuations are very sensitive to beam geometry. In particular, single scattering does not contribute to the total OAM fluctuations for beams with circular-symmetric irradiance distribution. As a result, beams that lack circular symmetry, for example, elliptical Gaussian beams, develop total OAM fluctuations much faster than Laguerre–Gaussian beams. It was also shown in [1] that the TLM and OAM densities can be measured using a Shack–Hartman wave front sensor.

In this report, we apply the ideas developed in [1] to investigate the statistics of the OAM of an optical wave intercepted by a finite aperture. By analogy with optical power flux fluctuations through an aperture, it can be called "OAM-in-the-bucket."

In Section, 2 we reiterate the relations between the OAM density and the coherence function of a propagating wave [1], and derive the principal equations for intrinsic aperture-averaged OAM. We recall the parabolic equations of the Markov Approximation theory [2] and examine the relationship of the mean value and variance of the aperture-averaged OAM with the second and fourth-order coherence functions of a propagating wave. It becomes clear that for the simple case of a spherical wave, the mean OAM maintains a zero value, but the OAM variance is affected by the fluctuations of the medium. An analytical solution for the OAM variance is not available and in Section 3, we perform a complete asymptotic analysis of this parameter in the simplest case of the Kolmogorov spectrum of the random medium, when the OAM variance depends only on two dimensionless parameters. Both weak and strong fluctuation cases are considered, and two-dimensional asymptotic maps are presented for circular-symmetric and non-circular symmetric apertures. In Section 4 we discuss the results of our asymptotic analysis and present several simple examples of the OAM variance dependence on aperture size and turbulence strength. A summary of the results and suggestions for further research are presented in Section 5. The appendix covers some technical aspects of the path-integral technique.

## 2. TLM AND OAM OF SCALAR PARAXIAL WAVES

### A. OAM and coherence function

For the case of a linear polarized narrowband paraxial wave propagating along the $z$-direction, when the scalar complex envelope of the electric field is $u(\mathbf{r}, z)$ the period-averaged

TLM density $\mathbf{L}(\mathbf{r},z)$ and the $z$-component of the OAM density $M(\mathbf{r},z)$ were calculated in [1, 9] as

$$\mathbf{L}(\mathbf{r},z) = \text{Im}\left[u^*(\mathbf{r},z)\nabla_r u(\mathbf{r},z)\right], M(\mathbf{r},z)\hat{\mathbf{z}} = \mathbf{r}\times\mathbf{L}(\mathbf{r},z). \quad (1)$$

Here, $\mathbf{r} = (x,y)$ is the transverse coordinate, and we dropped the inconsequential factor $\omega\varepsilon_0$. Both the TLM and OAM densities are related to the coherence function of the optical wave

$$\gamma(\mathbf{r},\boldsymbol{\rho},z) \equiv \overline{u\left(\mathbf{r}+\frac{\boldsymbol{\rho}}{2},z\right)u^*\left(\mathbf{r}-\frac{\boldsymbol{\rho}}{2},z\right)}. \quad (2)$$

The overbar in Eq. (2) indicates the averaging over the possible fluctuations of the radiation source. In the case of a coherent wave, the coherence function is just a product of the field and the conjugate of two points separated by a vector $\boldsymbol{\rho}$. It is straightforward to show [1, 3] that for the coherent case

$$\mathbf{L}(\mathbf{r},z) = -i\nabla_{\boldsymbol{\rho}}\gamma(\mathbf{r},0,z), M(\mathbf{r},z)\hat{\mathbf{z}} = -i\mathbf{r}\times\nabla_{\boldsymbol{\rho}}\gamma(\mathbf{r},0,z). \quad (3)$$

After presenting $u(\mathbf{r},z)$ in the amplitude and phase form and introducing the wave irradiance $I(\mathbf{r},z)$, we have

$$u(\mathbf{r},z) = A(\mathbf{r},z)\exp[i\varphi(\mathbf{r},z)],$$
$$A(\mathbf{r},z) \geq 0, I(\mathbf{r},z) = A^2(\mathbf{r},z) = \gamma(\mathbf{r},0,z), \quad (4)$$

the OAM density $M(\mathbf{r},z)$ and the total OAM of the beam wave $M(z)$ are

$$M(\mathbf{r},z)\hat{\mathbf{z}} = I(\mathbf{r},z)\mathbf{r}\times\nabla\varphi(\mathbf{r},z), M(z) = \iint d^2r M(\mathbf{r},z). \quad (5)$$

As discussed in [1, 3], the total TLM and OAM of the bounded beam waves are conserved in free-space propagation. In the turbulent propagation case, $M(z)$ fluctuates, but the mean values $\langle M(z)\rangle$ is conserved [1].

**B. Aperture-integrated intrinsic OAM**

Here, we consider the statistics of the OAM intercepted by a finite aperture for the simplest case of a propagating spherical wave with a point source located at $(\mathbf{0},L)$. After propagation over the turbulent path, the wave is incident on a finite aperture with a transmission function $A(\mathbf{r})$ located at $z=0$. The total OAM captured by the aperture is

$$M = \iint d^2r A(\mathbf{r})M(\mathbf{r}) = \iint d^2r A(\mathbf{r})\mathbf{r}\times\mathbf{L}(\mathbf{r})$$
$$= -i\iint d^2r A(\mathbf{r})\mathbf{r}\times\nabla_{\boldsymbol{\rho}}\gamma(\mathbf{r},0). \quad (6)$$

Here, and later on, we drop the notation for the $z$-dependence because all the wave parameters are considered in the same plane $z=0$. Based on the discussion in [1], this random parameter of an optical wave can be directly measured by a standard Shack–Hartman wave sensor without invoking any phase estimations. The first modeling and field measurements results of the aperture-averaged OAM were reported recently in [8].

It is well-known for the beam-wave case [10] and is clearly true for the aperture-integrated case that in general, the OAM depends on the reference point used for the position vector $\mathbf{r}$. Namely, if the reference point $\mathbf{r}_0$ is used instead of the coordinate origin, then

$$M_{\mathbf{r}_0} = -i\iint d^2r A(\mathbf{r})(\mathbf{r}-\mathbf{r}_0)\times\nabla_{\boldsymbol{\rho}}\gamma(\mathbf{r},0) = M - \mathbf{r}_0\times\mathbf{L}. \quad (7)$$

where

$$\mathbf{L} = \iint d^2r A(\mathbf{r})\mathbf{L}(\mathbf{r}) \quad (8)$$

is the aperture-integrated TLM. The intrinsic aperture-integrated OAM can be introduced if the average TLM density is removed from the TLM density $\mathbf{L}(\mathbf{r})$ in Eq. (6). Namely, the intrinsic aperture-integrated OAM is

$$\tilde{M} = \iint d^2r A(\mathbf{r})\mathbf{r}\times\left[\mathbf{L}(\mathbf{r})-\frac{\mathbf{L}}{S}\right], S \equiv \iint d^2r' A(\mathbf{r}'). \quad (9)$$

Using Eq. (7) in Eq. (9), it is straightforward to show that the intrinsic OAM can be presented as

$$\tilde{M} = \iint d^2r A(\mathbf{r})(\mathbf{r}-\mathbf{r}_A)\times\mathbf{L}(\mathbf{r}), \mathbf{r}_A \equiv \frac{1}{S}\iint d^2r' A(\mathbf{r}')\mathbf{r}'. \quad (10)$$

and corresponds to the use of the aperture centroid as a reference point for the OAM density definition. During further development, we assume that the coordinate origin is placed at the aperture centroid, and $\mathbf{r}_A = \mathbf{0}$. For the important class of circular-symmetric apertures, the $A(\mathbf{r}) = A(r)$ coordinate origin, naturally, is at the aperture center. One formal consequence of the placement of the origin at the centroid will affect our further development. If we introduce a normalized aperture function spectrum as

$$\hat{A}(\mathbf{p}) = \frac{1}{S}\iint d^2r A(\mathbf{r})e^{-i\mathbf{p}\cdot\mathbf{r}}, \quad (11)$$

then $\hat{A}(\mathbf{0}) = 1$ for a general coordinate choice, and additionally $\nabla\hat{A}(\mathbf{0}) = 0$ for the intrinsic OAM coordinates.

It is straightforward to show that for a simple single vortex field, we have

$$u(\mathbf{r}) = u(x,y) = u_0 x + i u_0 y, \quad (12)$$

And the corresponding coherence function, irradiance, and OAM density are

$$\gamma(\mathbf{r},\boldsymbol{\rho}) = u(x,y) = |u_0|^2\left(r^2 - \frac{1}{4}\rho^2 + i\mathbf{r}\times\boldsymbol{\rho}\right),$$
$$I(\mathbf{r}) = M(\mathbf{r}) = |u_0|^2 r^2 \quad (13)$$

The aperture-integrated OAM normalized by the power intercepted by aperture is

$$\frac{\widetilde{M}}{P}=1. \tag{14}$$

## C. OAM and turbulence propagation

Atmospheric turbulence along the propagation path $0 < z < L$ produces fluctuations of the OAM density $M(\mathbf{r})$ and aperture-averaged OAM $\widetilde{M}$. The statistical properties of these fluctuations can be described by the statistical moments averaged over the ensemble of turbulence samples along the propagation path. The mean OAM density is related to the second-order coherence function of a spherical wave

$$\Gamma(\mathbf{r},\boldsymbol{\rho}) \equiv \langle \gamma(\mathbf{r},\boldsymbol{\rho}) \rangle, \tag{15}$$

Here, averaging is performed over a statistical ensemble of turbulent fluctuations of the refractive index along the propagation path from the source to the aperture plane. According to Eq. (8) the mean aperture-integrated OAM is

$$\langle \widetilde{M} \rangle = -i \iint d^2 r A(\mathbf{r}) \mathbf{r} \times \nabla_{\boldsymbol{\rho}} \Gamma(\mathbf{r},0). \tag{16}$$

The propagation of the coherence function $\Gamma(\mathbf{r},\boldsymbol{\rho},z)$ is described by a well-known [2, 6] parabolic equation of the Markov approximation as

$$\frac{\partial \Gamma(\mathbf{r},\boldsymbol{\rho},z)}{\partial z} - \frac{i}{k}\nabla_r \cdot \nabla_\rho \Gamma(\mathbf{r},\boldsymbol{\rho},z) + \frac{\pi k^2}{4} H(\boldsymbol{\rho}) \Gamma(\mathbf{r},\boldsymbol{\rho},z) = 0. \tag{17}$$

Here,

$$H(\boldsymbol{\rho}) = 8 \iint d^2\kappa \, \Phi_n(\boldsymbol{\kappa},0)[1-\cos(\boldsymbol{\kappa}\boldsymbol{\rho})], \tag{18}$$

$k = 2\pi/\lambda$ is the wavenumber, and $\Phi_n(\boldsymbol{\kappa},p)$ is a 3-D spectrum of the refractive index fluctuations. The solution of Eq. (17) for the spherical wave is well known

$$\Gamma(\mathbf{r},\boldsymbol{\rho},L) = \frac{k^2}{4\pi^2 L^2} \exp\left( \frac{ik}{L} \mathbf{r}\cdot\boldsymbol{\rho} - \frac{\pi k^2}{4} \int_0^L dz H\left[\boldsymbol{\rho}\left(1-\frac{z}{L}\right)\right] \right) \tag{19}$$

and offers a straightforward path for calculation of the mean Irradiance, $\langle I(\mathbf{r}) \rangle$, TLM and OAM densities,

$$\langle I(\mathbf{r}) \rangle = \Gamma(\mathbf{r},0) = \frac{k^2}{4\pi^2 L^2}, \quad \langle \mathbf{L}(\mathbf{r}) \rangle = -i\nabla_\rho \Gamma(\mathbf{r},0) = \frac{k^2}{4\pi^2 L^2}\frac{k}{L}\mathbf{r}, \tag{20}$$
$$\langle M(\mathbf{r}) \rangle = \mathbf{r} \times \langle \mathbf{L}(\mathbf{r}) \rangle = 0.$$

This shows that the mean TLM density is not affected by turbulence and that the mean OAM density is conserved and maintains a zero value. As a consequence, the mean aperture averaged OAM, $\langle \widetilde{M} \rangle = 0$.

The simplest statistic that provides an insight into the OAM fluctuations is the normalized variance of the aperture-integrated intrinsic OAM which is given as

$$\sigma_{OAM}^2 = \frac{\langle \widetilde{M}^2 \rangle - \langle \widetilde{M} \rangle^2}{P^2}, \tag{21}$$

where $P$ is the mean power flux through the aperture. Because the OAM density has a dimensionality of irradiance, $\sigma_{OAM}$ is dimensionless. As suggested by Eq. (14), the numeric value of $\sigma_{OAM}$ can be loosely interpreted as the typical number of vortices that can be present at the aperture. We recall [1], that with similar normalization, the total normalized OAM of the Laguerre–Gaussian vortex beam is equal to the topological charge of the vortex.

Using Eqs. (6) and (20), the normalized variance of the aperture-integrated OAM can be presented as

$$\sigma_{OAM}^2 = -\left(\frac{4\pi^2 L^2}{Sk^2}\right)^2 \iint A(\mathbf{r}_1) d^2 r_1 \iint d^2 r_2 A(\mathbf{r}_2) \tag{22}$$
$$\times (\mathbf{r}_1 \times \nabla_{\boldsymbol{\rho}_1})(\mathbf{r}_2 \times \nabla_{\boldsymbol{\rho}_2}) \Gamma_4(\mathbf{r}_1,0,\mathbf{r}_2,0,z).$$

Here, the fourth-order coherence function is introduced as

$$\Gamma_4(\mathbf{r}_1,\boldsymbol{\rho}_1,\mathbf{r}_2,\boldsymbol{\rho}_2,z) = \langle \gamma(\mathbf{r}_1,\boldsymbol{\rho}_1,z)\gamma(\mathbf{r}_2,\boldsymbol{\rho}_2,z) \rangle, \tag{23}$$

and the second-order instantaneous coherence function is described by Eq. (2). The propagation equation for $\Gamma_4$ is well-known under the Markov approximation [2, 11]

$$\frac{\partial}{\partial z}\Gamma_4(\mathbf{r}_1,\boldsymbol{\rho}_1,\mathbf{r}_2,\boldsymbol{\rho}_2,z) - \frac{i}{k}\left(\nabla_{r_1}\cdot\nabla_{\rho_1} + \nabla_{r_2}\cdot\nabla_{\rho_2}\right)\Gamma_4$$
$$+ \frac{\pi k^2}{4}\Psi(\mathbf{r}_1,\boldsymbol{\rho}_1,\mathbf{r}_2,\boldsymbol{\rho}_2)\Gamma_4 = 0, \tag{24}$$

where the scattering potential is given by

$$\Psi(\mathbf{r}_1,\boldsymbol{\rho}_1,\mathbf{r}_2,\boldsymbol{\rho}_2) = H(\boldsymbol{\rho}_1) + H(\boldsymbol{\rho}_2)$$
$$+ H\left(\mathbf{r}_1 - \mathbf{r}_2 + \frac{\boldsymbol{\rho}_1+\boldsymbol{\rho}_2}{2}\right) + H\left(\mathbf{r}_1 - \mathbf{r}_2 - \frac{\boldsymbol{\rho}_1+\boldsymbol{\rho}_2}{2}\right) \tag{25}$$
$$- H\left(\mathbf{r}_1 - \mathbf{r}_2 + \frac{\boldsymbol{\rho}_1-\boldsymbol{\rho}_2}{2}\right) - H\left(\mathbf{r}_1 - \mathbf{r}_2 - \frac{\boldsymbol{\rho}_1-\boldsymbol{\rho}_2}{2}\right).$$

Equation (24) does not have an analytic solution. For further investigation, we resort to the asymptotic analysis of $\sigma_{OAM}^2$.

## 3. ASYMPTOTIC SOLUTIONS FOR $\sigma_{OAM}^2$.

In the case of Kolmogorov turbulence along a horizontal path

$$\Phi_n(\mathbf{K}) = 0.033 C_n^2 K^{-11/3}, \quad H(\mathbf{r}) = c_H C_n^2 r^{5/3}, \quad c_H = 1.86. \tag{26}$$

For the point source case after normalizing the longitudinal variable by the path length $L$, and transverse coordinates by the Fresnel radius $\sqrt{L/k}$, it is possible to present Eq. (24) in dimensionless form and to show that it depends on a single dimensionless parameter. Traditionally, this parameter is chosen to be the so-called "Rytov variance"

$$\sigma_I^2 = 0.50 C_n^2 k^{7/6} L^{11/6}. \tag{27}$$

However, for the purpose of our analysis it is more convenient to use the alternative parameter: Fresnel number,

which corresponds to the coherence radius $r_C$ of a propagating wave as

$$q \equiv \frac{kr_C^2}{L}, \quad r_C = (1.46 C_n^2 k^2 L)^{-3/5}, \quad q = 0.347(\sigma_I^2)^{-6/5}, \quad (28)$$

Each of these parameters completely determine the four point statistics of a spherical wave for Kolmogorov turbulence, Eq. (26). However, calculation of $\sigma_{OAM}^2$ introduces an aperture function $A(\mathbf{r})$, which we assume has an effective width $a$. This requires the introduction of an additional dimensionless parameter: aperture Fresnel number

$$N \equiv \frac{ka^2}{L}. \quad (29)$$

The normalized variance of the aperture-integrated OAM is a function of the two dimensionless parameters $q$ and $N$. In this section, we investigate this dependence using an asymptotic analysis of the function $\sigma_{OAM}^2(N,q)$. The general approach is similar to the one used in [12, 13], where asymptotes of a Gaussian beam scintillation index and the aperture averaged power flux variance were calculated.

Equation (24) does not have a closed-form analytic solution, but the solution can be presented in the form of a Feynman path integral, [7, 14], Eq. (A1). We use this path-integral representation to examine $\sigma_{OAM}^2(N,q)$ in two limiting cases of weak or strong fluctuations. This asymptotic technique was described in more details elsewhere [15, 13, 14], and only a brief description is given in the Appendix. The following two subsections discuss the weak and strong scintillation cases for a general turbulence spectrum and in the last subsection, we build a complete map of $\sigma_{OAM}^2(N,q)$ asymptotes for the Kolmogorov spectrum case.

**A. Weak Fluctuations**

The weak fluctuation case is described by the Perturbation Theory (PT) solution of the parabolic Eq. (24) or the equivalent path-integral, Eq. (A1) with respect to the scattering potential $\Psi$. The zero-order PT solution can be derived either by setting the scattering potential $\Psi \equiv 0$ either in parabolic Eq. (24), or in the path-integral, Eq. (A1). The zero-order PT term is just a free-space paraxial spherical wave given as

$$\Gamma_4^{(0)}(\mathbf{r}_1, \boldsymbol{\rho}_1, \mathbf{r}_2, \boldsymbol{\rho}_2) = \left(\frac{k^2}{4\pi^2 L^2}\right)^2 \exp\left(\frac{ik}{L}(\mathbf{r}_1 \cdot \boldsymbol{\rho}_1 + \mathbf{r}_2 \cdot \boldsymbol{\rho}_2)\right) \quad (30)$$

and it makes no contribution to $\sigma_{OAM}^2$.

The first- and second-order PT solutions of Eq. (24) can be derived by keeping the linear and quadratic terms of a Taylor series, Eq. (A3) in the path integral, Eq. (A1). After using the spectral expansion, Eq. (18), the path integrals can be calculated analytically, as described in [12], and the first-order term is

$$\Gamma_4^{(1)}(\mathbf{r}_1, \boldsymbol{\rho}_1, \mathbf{r}_2, \boldsymbol{\rho}_2) = \Gamma_2(\mathbf{r}_1, \boldsymbol{\rho}_1)\Gamma_2(\mathbf{r}_2, \boldsymbol{\rho}_2) + \Gamma_4^{(0)}(\mathbf{r}_1, \boldsymbol{\rho}_1, \mathbf{r}_2, \boldsymbol{\rho}_2) F(\mathbf{r}_1, \boldsymbol{\rho}_1, \mathbf{r}_2, \boldsymbol{\rho}_2) \quad (31)$$

where

$$F(\mathbf{r}_1, \boldsymbol{\rho}_1, \mathbf{r}_2, \boldsymbol{\rho}_2) \equiv 4\pi k^2 \int_0^L dz \iint d^2\kappa \, \Phi_n(\boldsymbol{\kappa}) \exp[i\boldsymbol{\kappa} \cdot (\mathbf{r}_1 - \mathbf{r}_2)]$$
$$\times \left\{ \cos\left[\boldsymbol{\kappa} \cdot \frac{(\boldsymbol{\rho}_1 + \boldsymbol{\rho}_2)}{2}\left(1 - \frac{z}{L}\right)\right] - \cos\left[\boldsymbol{\kappa} \cdot \frac{(\boldsymbol{\rho}_1 - \boldsymbol{\rho}_2)}{2}\left(1 - \frac{z}{L}\right) - \frac{\kappa^2 z}{k}\left(1 - \frac{z}{L}\right)\right] \right\}. \quad (32)$$

In the case of four collocated points $F(\mathbf{r}, 0, \mathbf{r}, 0)$ produces the PT result for the scintillation index as

$$\sigma_I^2 = \frac{\langle I^2 \rangle}{\langle I \rangle^2} - 1 = \frac{\Gamma_4^{(1)}(\mathbf{r}, 0, \mathbf{r}, 0)}{[\Gamma_2(\mathbf{r}, 0)]^2} - 1$$
$$= 4\pi k^2 \int_0^L dz \iint d^2\kappa \, \Phi_n(\boldsymbol{\kappa}) \exp[i\boldsymbol{\kappa} \cdot (\mathbf{r}_1 - \mathbf{r}_2)], \quad (33)$$
$$\times \left\{1 + \cos\left[\frac{\kappa^2 z}{k}\left(1 - \frac{z}{L}\right)\right]\right\}$$

which leads to the "Rytov variance," Eq. (27) in the case of Kolmogorov turbulence on a horizontal path Eq. (26).

The normalized variance of the OAM density in the first PT order can be calculated from the fourth-order coherence function, Eq. (31), as follows

$$\frac{\langle M^2(\mathbf{r}) \rangle}{\langle I \rangle^2} = -(\mathbf{r} \times \nabla_{\boldsymbol{\rho}_1})(\mathbf{r} \times \nabla_{\boldsymbol{\rho}_2}) F(\mathbf{r}, 0, \mathbf{r}, 0) = \pi k^2 \int_0^L dz \left(1 - \frac{z}{L}\right)^2$$
$$\times \iint d^2\kappa \, \Phi_n(\boldsymbol{\kappa})(\boldsymbol{\kappa} \times \mathbf{r})^2 \left\{1 + \cos\left[\frac{\kappa^2 z}{k}\left(1 - \frac{z}{L}\right)\right]\right\}. \quad (34)$$

It is noteworthy that the OAM density is an inhomogeneous, zero-mean, random field that has a zero variance at the origin, with a variance which grows as the square of the distance from the origin. For the Kolmogorov spectrum, Eq. (26), the spectral integral in Eq. (34) diverges at high-wavenumbers. This indicates that the OAM density variance is sensitive to the inner scale of the turbulence.

Using Eq. (22) and (31) we can calculate the variance of the aperture-averaged OAM $\sigma_{OAM}^2$ in the first PT order. We use the notation $\sigma_{PT1}^2$ to explicitly specify the approximation used as

$$\sigma_{PT1}^2 = \pi k^2 \int_0^L dz \left(1 - \frac{z}{L}\right)^2 \iint d^2\kappa \, \Phi_n(\boldsymbol{\kappa}) \left\{1 + \cos\left[\frac{\kappa^2 z}{k}\left(1 - \frac{z}{L}\right)\right]\right\}$$
$$\times \left|\frac{1}{S} \iint d^2 r A(\mathbf{r})(\boldsymbol{\kappa} \times \mathbf{r}) \exp\left[i\boldsymbol{\kappa} \cdot \mathbf{r}\left(1 - \frac{z}{L}\right)\right]\right|^2$$
. (35)

Using the normalized aperture function spectrum, Eq. (11), $\sigma_{PT1}^2$ can be presented as

$$\sigma_{PT1}^2 = \pi k^2 \int_0^L dz \left(1 - \frac{z}{L}\right)^2 \iint d^2\kappa \Phi_n(\kappa) \tag{36}$$
$$\times \left\{1 + \cos\left[\frac{\kappa^2 z}{k}\left(1 - \frac{z}{L}\right)\right]\right\} \left|\kappa \times \nabla \hat{A}\left[\kappa\left(1 - \frac{z}{L}\right)\right]\right|^2$$

It is easy to show that for any circular-symmetric aperture, we have

$$A(\mathbf{r}) = A(r), \ \hat{A}(\mathbf{K}) = \hat{A}(K), \ \nabla \hat{A}(\mathbf{K}) = \hat{B}(K)\mathbf{K}, \ \hat{B}(K) = \frac{\hat{A}'(K)}{K}, \tag{37}$$

and $\sigma_{PT1}^2 = 0$. Thus, the first-order PT term makes no contribution to the aperture-averaged OAM fluctuations. This statement is valid for any circular-symmetric aperture, and any turbulence spectrum including anisotropic turbulence. However, there will be a non-vanishing first-order term, for example, for an elliptical or a semi-circular aperture. The situation here is similar to the one encountered for the total OAM of bounded beam waves discussed in [1], where the total OAM variance for elliptical Gaussian beams was calculated. The order-of-magnitude estimates of $\sigma_{PT1}^2$ for the Kolmogorov turbulence spectrum, expressed by Eq. (26), will be presented at the end of this Section.

The vanishing of $\sigma_{PT1}^2$ suggests that the leading PT term for the aperture-averaged OAM variance should be sought in the second-order term of the perturbation series. We calculated the second-order PT term $\Gamma_4^{(2)}(\mathbf{r}_1, \boldsymbol{\rho}_1, \mathbf{r}_2, \boldsymbol{\rho}_2)$ using the Feynman path-integral solution of the parabolic Eq. (24), but we will not show it here due to its considerable size. The second-order PT result for $\sigma_{OAM}^2$, which we designate as $\sigma_{PT2}^2$ is more manageable, and we present the results for the circular-symmetric aperture case

$$\sigma_{PT2}^2 = 4(\pi k^2)^2 \int_0^L dz_1 \left(1 - \frac{z_1}{L}\right)^2 \int_0^L dz_2 \left(1 - \frac{z_2}{L}\right)^2 \iint d^2\kappa_1 \Phi_n(\kappa_1)$$
$$\times \iint d^2\kappa_1 \Phi_n(\kappa_2)(\kappa_1 \times \kappa_2)^2 \hat{B}^2\left(\left|\kappa_1\left(1 - \frac{z_1}{L}\right) + \kappa_2\left(1 - \frac{z_2}{L}\right)\right|\right) \tag{38}$$
$$\times \left\{1 - \cos\left[\frac{\kappa_1^2 z_1}{k}\left(1 - \frac{z_1}{L}\right) - \frac{\kappa_2^2 z_2}{k}\left(1 - \frac{z_2}{L}\right)\right]\right\}.$$

The auxiliary function $\hat{B}(K)$, was introduced in Eq. (37). Eq. (38) is still fairly complicated, but it is possible to examine the two asymptotic cases. It is easy to see that for $N < 1$

$$\sigma_{PT2}^2 = 4(\pi k^2)^2 \int_0^L dz_1 \left(1 - \frac{z_1}{L}\right)^2 \int_0^L dz_2 \left(1 - \frac{z_2}{L}\right)^2 \iint d^2\kappa_1 \Phi_n(\kappa_1)$$
$$\times \iint d^2\kappa_1 \Phi_n(\kappa_2)(\kappa_1 \times \kappa_2)^2 B^2\left(\left|\kappa_1\left(1 - \frac{z_1}{L}\right) + \kappa_2\left(1 - \frac{z_2}{L}\right)\right|\right), \tag{39}$$

and for $N > 1$

$$\sigma_{PT2}^2 = 2(\pi k^2)^2 \int_0^L dz_1 \left(1 - \frac{z_1}{L}\right)^2 \int_0^L dz_2 \left(1 - \frac{z_2}{L}\right)^2 \iint d^2\kappa_1 \Phi_n(\kappa_1)$$
$$\times \iint d^2\kappa_1 \Phi_n(\kappa_2) \hat{B}^2\left(\left|\kappa_1\left(1 - \frac{z_1}{L}\right) + \kappa_2\left(1 - \frac{z_2}{L}\right)\right|\right) \tag{40}$$
$$\times (\kappa_1 \times \kappa_2)^2 \left[\kappa_1^2 z_1 \left(1 - \frac{z_1}{L}\right) - \kappa_2^2 z_2 \left(1 - \frac{z_2}{L}\right)\right]^2$$

The calculation of these multiple integrals is still a difficult task, but they allow us to provide the parameterization of $\sigma_{OAM}^2$, at least for the Kolmogorov spectrum of turbulence, Eq. (26), which is shown at the end of this Section.

**B. Strong Fluctuations**

In this section, we use the CC expansion described in the Appendix to calculate the leading asymptotic terms that are valid under the strong fluctuation conditions. These calculations use spectral expansion, Eq. (18) and certain path integral identities. We refer to [12, 13, 16] for details.

It is easy to show that the density of the OAM variance which corresponds to the zero-order MCC term, Eq. (A6), is zero at any point on the wave front and does not contribute to $\sigma_{OAM}^2$. The contribution of the zero-order ACC term, Eq. (A7), to $\sigma_{OAM}^2$, which we denote as $\sigma_{ACC0}^2$, can be calculated from Eq. (22) as

$$\sigma_{ACC0}^2 = \frac{\pi k^2}{8S^2} \iint A^2(\mathbf{R}) d^2 R \int_0^L dz \left(1 - \frac{z}{L}\right)^2 \iint d^2 r$$
$$\times \exp\left[-\frac{\pi k^2}{2} \int_0^L dz' H\left(\mathbf{r}\left(1 - \frac{z'}{L}\right)\right)\right] (\mathbf{R} \times \nabla)^2 H\left(\mathbf{r}\left(1 - \frac{z'}{L}\right)\right). \tag{41}$$

This equation can be further simplified for isotropic (in the plane orthogonal to the propagation direction) turbulence when

$$H(\mathbf{r}) = H(r), \ \nabla H(\mathbf{r}) = G(r)\mathbf{r}, \ G(r) = \frac{H'(r)}{r} \tag{42}$$

by integrating over the azimuthal angle

$$\sigma_{ACC0}^2 = \left[\frac{\iint R^2 A^2(\mathbf{R}) d^2 R}{\left(\iint A(\mathbf{R}) d^2 R\right)^2}\right] \frac{\pi^2 k^2}{8} \int_0^L dz \left(1 - \frac{z}{L}\right)^2$$
$$\times \int_0^\infty r dr \exp\left[-\frac{\pi k^2}{2} \int_0^L d\zeta H\left(r\left(1 - \frac{\zeta}{L}\right)\right)\right] \tag{43}$$
$$\times \left[H''\left(r\left(1 - \frac{z}{L}\right)\right) + G\left(r\left(1 - \frac{z}{L}\right)\right)\right],$$

where the auxiliary function $G(r)$ was introduced in Eq. (42). The order-of-magnitude estimates for $\sigma_{ACC0}^2$ will be presented in the next subsection.

As was mentioned earlier, the zero-order MCC term makes no contribution to $\sigma^2_{OAM}$. It is therefore necessary to consider the first-order term of the MCC series. Substitution of Eq. (A8) into Eq. (22) results in

$$\sigma^2_{MCC1} = 4\pi k^2 \int_0^L dz \iint d^2\kappa \Phi_n(\kappa) \exp\left[-\frac{\pi k^2}{2}\int_0^L d\zeta H(\mathbf{s}(z,\zeta))\right]$$
$$\times \left\{[1-\cos(\kappa \cdot \mathbf{s}(z,z))]\right.$$
$$\times \left|\frac{\pi k^2}{4}\int_0^L dz'\left(1-\frac{z'}{L}\right)\nabla \hat{A}\left(\kappa\left(1-\frac{z}{L}\right)\right)\times \nabla H(\mathbf{s}(z,z'))\right|^2$$
$$\left. + \frac{1}{4}\left(1-\frac{z}{L}\right)^2[1+\cos(\kappa \cdot \mathbf{s}(z,z))]\left|\kappa \times \nabla \hat{A}\left(\kappa\left(1-\frac{z}{L}\right)\right)\right|^2\right\}.$$
(44)

Here,
$$\mathbf{s}(z,\zeta) = \frac{\kappa}{k}\min(z,\zeta)\left(1-\frac{\max(z,\zeta)}{L}\right) \quad (45)$$

is a set of single-kink paths that contribute to the path-integral in this case. In the simple, but practically most important case of isotropic turbulence and circular-symmetric apertures, both $\nabla \hat{A}$, and $\nabla H$ in Eq. (43) are collinear with $\kappa$, Eqs. (37) and (42), and $\sigma^2_{MCC1} = 0$. This is somewhat similar to the first-order perturbation theory result discussed in the previous subsection, with the additional requirement that the coherence channel determined by $H(\mathbf{s}(z,\zeta))$ is also isotropic.

The general case of non-rotation-symmetric apertures requires a more detailed analysis of Eq. (44). As is mentioned in the Appendix, the validity of the CC expansion requires that $|\mathbf{s}(z,\zeta)| \propto r_C$, or $\kappa \propto kr_C/L$. Assuming, as is sufficient for our further investigation, that the turbulence spectrum is fractal and imposes no scale restrictions, the only alternative scaling factor is the aperture spectrum $\hat{A}$ where $\kappa \propto 1/a$. Hence, in order to have a valid CC expansion, the condition $qN < 1$ needs to be imposed, and the small-argument assumption must be used for $\nabla \hat{A}$. Recalling that as was discussed in Subsection 2.2, $\nabla \hat{A}(\mathbf{0}) = 0$ for intrinsic OAM, we have the following Cartesian coordinate representation of $\nabla \hat{A}$ for small arguments

$$\nabla \hat{A}\left(\kappa\left(1-\frac{z}{L}\right)\right) \approx \left(1-\frac{z}{L}\right)\begin{pmatrix}\hat{A}_{xx}(0)\kappa_x + \hat{A}_{xy}(0)\kappa_y \\ \hat{A}_{xy}(0)\kappa_x + \hat{A}_{yy}(0)\kappa_y\end{pmatrix}. \quad (46)$$

Here, $\nabla \hat{A}$ is presented as a column vector. We limit our scope by isotropic turbulence case when Eq. (44), after the application of Eq. (46), and the integration over the angular variable simplifies as follows

$$\sigma^2_{MCC1} = \pi^2 k^2 \left(K_1^2 - K_2^2\right)^2 \int_0^L dz\left(1-\frac{z}{L}\right)^2 \int_0^\infty d\kappa \kappa^5 \Phi_n(\kappa)$$
$$\times \exp\left[-\frac{\pi k^2}{2}\int_0^L d\zeta H(s(z,\zeta))\right]$$
$$\times \left\{\frac{[1-\cos(\kappa s(z,z))]}{\kappa^2}\left|\frac{\pi k^2}{4}\int_0^L dz'\left(1-\frac{z'}{L}\right)H'(s(z,z'))\right|^2\right.$$
$$\left. + \frac{1}{4}\left(1-\frac{z}{L}\right)^2[1+\cos(\kappa s(z,z))]\right\}.$$
(47)

Here,
$$\left(K_1^2 - K_2^2\right)^2 = \hat{A}_{xx}^2(0) + \hat{A}_{yy}^2(0) - 2\hat{A}_{xx}(0)\hat{A}_{yy}(0) + 4\hat{A}_{xy}^2(0) \quad (48)$$

is the squared difference of the principal curvatures of $\hat{A}(\mathbf{p})$ at zero frequency, which can also be presented in terms of the second geometrical moments of the aperture function $A(\mathbf{r})$. Eq. (47) confirms our earlier observation that $\sigma^2_{MCC1} = 0$ for rotationally-symmetrical apertures where the principal curvatures are equal.

It is not necessary to elaborate on the details of the first-order ACC term here, as the zero-order term $\sigma^2_{ACC0}$, Eq. (43) is not zero, and $\sigma^2_{ACC1}$ can only be a small correction to $\sigma^2_{ACC0}$ in the validity domain of the CC expansion. However, an analysis of the integral representation of $\sigma^2_{ACC1}$, not shown here, reveals that the necessary condition for the validity of the ACC term is $q < N$. Combined with the earlier MCC-based constraint $qN < 1$, we can present the validity domain of the CC expansion as

$$q < \min(N, N^{-1}). \quad (49)$$

In particular, this implies that $q < 1$ for both the MCC and ACC terms, and Eq. (47) can be further simplified as

$$\sigma^2_{MCC1} = \frac{\pi^2 k^2}{2}\left(K_1^2 - K_2^2\right)^2 \int_0^L dz\left(1-\frac{z}{L}\right)^4 \int_0^\infty d\kappa \kappa^5$$
$$\times \Phi_n(\kappa)\exp\left[-\frac{\pi k^2}{2}\int_0^L d\zeta H(s(z,\zeta))\right]$$
$$\times \left[\frac{\kappa^2 z^2}{k^2}\left|\frac{\pi k^2}{4}\int_0^L dz'\left(1-\frac{z'}{L}\right)H'(s(z,z'))\right|^2 + 1\right].$$
(50)

The order-of-magnitude estimates of $\sigma^2_{MCC1}$ for the Kolmogorov turbulence case will be presented in the following subsection.

Because neither the zero nor the first term of the MCC series contribute to $\sigma_{OAM}^2$ in the practically important case of circular symmetric apertures, it is necessary to consider the contribution of the second-order MCC term. After substitution of Eq. (A10) into Eq. (22) and using the normalized aperture spectrum Eq. (11), we have

$$\sigma_{MCC2}^2 = 2\pi^2 k^4 \int_0^L dz_1 \int_0^L dz_2 \iint d^2\kappa_1 \Phi_n(\kappa_1)$$
$$\times \iint d^2\kappa_2 \Phi_n(\kappa_2) \exp\left\{-\frac{\pi k^2}{2}\int_0^L d\zeta H[\mathbf{t}(z_1,z_2,\zeta)]\right\} \quad (51)$$
$$\times \left(\mathbf{P}_1 \times \nabla A^*\left[\kappa_1\left(1-\frac{z_1}{L}\right)+\kappa_2\left(1-\frac{z_2}{L}\right)\right]\right)$$
$$\times \left(\mathbf{P}_2 \times \nabla A\left[\kappa_1\left(1-\frac{z_1}{L}\right)+\kappa_2\left(1-\frac{z_2}{L}\right)\right]\right).$$

Here, we used shorthand notations

$$\mathbf{P}_1 = \frac{\pi k^2}{4}\int_0^L d\eta\left(1-\frac{\eta}{L}\right)\nabla H[-\mathbf{t}(z_1,z_2,\eta)]$$
$$\times \sin[\kappa_1 \cdot \mathbf{t}(z_1,z_2,z_1)]\sin[\kappa_2 \cdot \mathbf{t}(z_1,z_2,z_2)]$$
$$+ \kappa_1\left(1-\frac{z_1}{L}\right)\cos[\kappa_1 \cdot \mathbf{t}(z_1,z_2,z_1)]\sin[\kappa_2 \cdot \mathbf{t}(z_1,z_2,z_2)] \quad (52)$$
$$+ \kappa_2\left(1-\frac{z_2}{L}\right)\cos[\kappa_2 \cdot \mathbf{t}(z_1,z_2,z_2)]\sin[\kappa_1 \cdot \mathbf{t}(z_1,z_2,z_1)],$$

$$\mathbf{P}_2 = \frac{\pi k^2}{4}\int_0^L d\eta\left(1-\frac{\eta}{L}\right)\nabla H[\mathbf{t}(z_1,z_2,\eta)]$$
$$\times \sin[\kappa_1 \cdot \mathbf{t}(z_1,z_2,z_1)]\sin[\kappa_2 \cdot \mathbf{t}(z_1,z_2,z_2)]$$
$$- \kappa_1\left(1-\frac{z_1}{L}\right)\cos[\kappa_1 \cdot \mathbf{t}(z_1,z_2,z_1)]\sin[\kappa_2 \cdot \mathbf{t}(z_1,z_2,z_2)] \quad (53)$$
$$- \kappa_2\left(1-\frac{z_2}{L}\right)\cos[\kappa_2 \cdot \mathbf{t}(z_1,z_2,z_2)]\sin[\kappa_1 \cdot \mathbf{t}(z_1,z_2,z_1)],$$

and

$$\mathbf{t}(z_1,z_2,\zeta) = \frac{\kappa_1}{k}\min(z_1,\zeta)\left[1-\frac{\max(z_1,\zeta)}{L}\right]$$
$$+ \frac{\kappa_2}{k}\min(z_2,\zeta)\left[1-\frac{\max(z_2,\zeta)}{L}\right]. \quad (54)$$

We are interested in the second-order MCC term mostly for isotropic turbulence and the circular-symmetric aperture case when the zero and first-order terms vanish. Using the straightforward extension of Eq. (46), we rewrite Eq. (51) as

$$\sigma_{MCC2}^2 = 32\pi^2 k^4 \int_0^L dz_1 \int_0^L dz_2 \iint d^2\kappa_1 \Phi_n(\kappa_1) \iint d^2\kappa_2 \Phi_n(\kappa_2)$$
$$\times [\kappa_1 \times \kappa_2]^2 \hat{B}^2\left[\left|\kappa_1\left(1-\frac{z_1}{L}\right)+\kappa_2\left(1-\frac{z_2}{L}\right)\right|\right] \quad (55)$$
$$\times \exp\left\{-\frac{\pi k^2}{2}\int_0^L d\zeta H[t(z_1,z_2,\zeta)]\right\} V^2(z_1,z_2,\kappa_1,\kappa_2).$$

Here,

$$V(z_1,z_2,\kappa_1,\kappa_2) = \frac{\pi k}{4}\int_0^L d\eta\left(1-\frac{\eta}{L}\right) G[\mathbf{t}(z_1,z_2,\eta)]$$
$$\times \left[\left(1-\frac{z_2}{L}\right)\min(z_1,\eta)\left(1-\frac{\max(z_1,\eta)}{L}\right)\right.$$
$$\left.-\left(1-\frac{z_1}{L}\right)\min(z_2,\eta)\left(1-\frac{\max(z_2,\eta)}{L}\right)\right] \quad (56)$$
$$\times \sin\left[\frac{\kappa_1 \cdot \mathbf{t}(z_1,z_2,z_1)}{2}\right]\sin\left[\frac{\kappa_2 \cdot \mathbf{t}(z_1,z_2,z_2)}{2}\right]$$
$$+ \frac{1}{2}\left(1-\frac{z_1}{L}\right)\left(1-\frac{z_2}{L}\right)\sin\left[\frac{\kappa_1^2 z_1}{2k}\left(1-\frac{z_1}{L}\right)-\frac{\kappa_2^2 z_2}{2k}\left(1-\frac{z_2}{L}\right)\right].$$

Recalling the CC constraint Eq. (49), we can further simplified Eq. (55) as

$$\sigma_{MCC2}^2 = 2\pi^2 k^4 \hat{A}_{xx}^2(0)\int_0^L dz_1 \int_0^L dz_2 \iint d^2\kappa_1 \Phi_n(\kappa_1) \iint d^2\kappa_2 \Phi_n(\kappa_2) \quad (57)$$
$$\times [\kappa_1 \times \kappa_2]^2 \exp\left\{-\frac{\pi k^2}{2}\int_0^L d\zeta H[t(z_1,z_2,\zeta)]\right\} W^2(z_1,z_2,\kappa_1,\kappa_2).$$

where

$$W(z_1,z_2,\kappa_1,\kappa_2) = \frac{\pi k}{4}\int_0^L d\eta\left(1-\frac{\eta}{L}\right) G[\mathbf{t}(z_1,z_2,\eta)]$$
$$\times \left[\left(1-\frac{z_2}{L}\right)\min(z_1,\eta)\left(1-\frac{\max(z_1,\eta)}{L}\right)\right.$$
$$\left.-\left(1-\frac{z_1}{L}\right)\min(z_2,\eta)\left(1-\frac{\max(z_2,\eta)}{L}\right)\right] \quad (58)$$
$$\times [\kappa_1 \cdot \mathbf{t}(z_1,z_2,z_1)][\kappa_2 \cdot \mathbf{t}(z_1,z_2,z_2)]$$
$$+ \left(1-\frac{z_1}{L}\right)\left(1-\frac{z_2}{L}\right)\left[\frac{\kappa_1^2 z_1}{k}\left(1-\frac{z_1}{L}\right)-\frac{\kappa_2^2 z_2}{k}\left(1-\frac{z_2}{L}\right)\right].$$

Here, for a circular-symmetric aperture, the OAM reference point is the center of the aperture, and in Eq. (46) $A_{yy}(0) = A_{xx}(0)$, and $A_{xy}(0) = 0$.

### C. Asymptotic Map of $\sigma_{OAM}^2$ for Kolmogorov Turbulence

In Subsection 3.2, several expressions for $\sigma_{OAM}^2$ that are based on either the PT, Eqs. (36), (39), (40), or on the CC expansion,

Eqs. (43), (47), and (57) were presented. While it is intuitively clear that the PT is useable for weak turbulence, and the CC expansion should be used for very strong turbulence, the rigorous determination of the validity domains of the PT and CC expansions is a challenging problem. We refer to [12, 13, 16] for several examples of this analysis, and to [17] for a short review that shows that the weak/strong turbulence boundary is problem-specific, and typically, is more complicated than the ubiquitous condition $\sigma_I^2 \approx 1$.

In this case, we limit ourselves to the simplest case of Kolmogorov turbulence Eq. (26) when $\sigma_{OAM}^2$ depends only on the two parameters $N$ and $q$ - the Fresnel numbers for the aperture size and coherence radius, Eq. (28) and (29). Strong inequalities defining the validity conditions of the different asymptotic solutions for $\sigma_{OAM}^2$ determine certain domains at the parameter plane $(N, q)$. A combination of these domains form an asymptotic map covering the $(N, q)$ plane. Inside each domain, a certain asymptotic solution for $\sigma_{OAM}^2$ is valid. Two asymptotes residing at adjacent domains has to be of the same order of magnitude at the boundary between these domains.

### 1. Non-circular-symmetric apertures

For non-rotationally-symmetric apertures, the first-order PT solution given by Eq. (36) for turbulence spectrum, Eq. (26) is

$$\sigma_{PT1}^2 = 0.033\pi k^2 \int_0^L dz C_n^2 \left(1 - \frac{z}{L}\right)^2 \iint d^2\kappa \kappa^{-11/3}$$

$$\times \left\{1 + \cos\left[\frac{\kappa^2 z}{k}\left(1 - \frac{z}{L}\right)\right]\right\} \left|\kappa \times \nabla \hat{A}\left[\kappa\left(1 - \frac{z}{L}\right)\right]\right|^2 \quad (59)$$

Assuming that a non-circular-symmetric aperture function $A(\mathbf{r})$ can be characterized by a single scale $a$, we infer that $\kappa \propto a^{-1}$ and $\sigma_{PT1}^2$ has two asymptotic cases depending on the value of the Fresnel number $N$. For $N < 1$, the cosine function in Eq. (57) oscillate rapidly in the effective integration area $\kappa \propto a^{-1}$. The corresponding term under the integral makes a small contribution to $\sigma_{PT1}^2$ and we have

$$\sigma_{PT1}^2 = 0.033\pi k^2 \int_0^L dz C_n^2 \left(1 - \frac{z}{L}\right)^2 \iint d^2\kappa \kappa^{-11/3} \left|\kappa \times \nabla \hat{A}\left[\kappa\left(1 - \frac{z}{L}\right)\right]\right|^2 \quad (60)$$

$$= O\left(C_n^2 k^2 L a^{5/3}\right) = O\left(r_C^{-5/3} a^{5/3}\right) = O\left(q^{-5/6} N^{5/6}\right).$$

In the opposite case $N > 1$, the cosine function argument is small, and we have

$$\sigma_{PT1}^2 = 2 \cdot 0.033\pi k^2 \int_0^L dz C_n^2 \left(1 - \frac{z}{L}\right)^2 \iint d^2\kappa \kappa^{-11/3} \left|\kappa \times \nabla \hat{A}\left[\kappa\left(1 - \frac{z}{L}\right)\right]\right|^2, (61)$$

$$= O\left(C_n^2 k^2 L a^{5/3}\right) = O\left(r_C^{-5/3} a^{5/3}\right) = O\left(q^{-5/6} N^{5/6}\right)$$

In both cases $\sigma_{PT1}^2$ has the same order of magnitude in terms of the dimensionless parameters $q$ and $N$, but the numeric coefficient doubles in the transition from small to large aperture Fresnel numbers.

The CC solution for non-circular-symmetric apertures is presented as a sum on the $\sigma_{ACC0}^2$, Eq. (43) and $\sigma_{MCC1}^2$, Eq. (50). The integrals in Eq. (43) can be explicitly calculated for $C_n^2 = const$, and $\sigma_{ACC0}^2$ is given by

$$\sigma_{ACC0}^2 = \frac{5\pi}{12} \left[\frac{\iint R^2 A^2(\mathbf{R}) d^2 R}{\left(\iint A(\mathbf{R}) d^2 R\right)^2}\right] = O(1). \quad (62)$$

In Eq. (50), $\kappa \propto L/kr_C$ and the order-of-magnitude of $\sigma_{MCC1}^2$ is estimated as

$$\sigma_{MCC1}^2 = O\left(r_C^{2/3} a^4 k^{7/3} L^{-7/3}\right) = O\left(q^{1/3} N^2\right). \quad (63)$$

Here, we assumed that the second geometrical moments of $A(\mathbf{r})$ are $\propto a^2$.

Finally, the estimate of $\sigma_{OAM}^2$ in the validity domain of the CC expansion is

$$\sigma_{OAM}^2 = \sigma_{ACC0}^2 + \sigma_{MCC1}^2 = O(1) + O\left(q^{1/3} N^2\right). \quad (64)$$

From the derivation of the terms of the CC expansion, we should expect that Eq. (64) is valid in the domain given by Eq. (49), which is illustrated as an equilateral triangle at the lower part of Fig. 1. In this domain, the $\sigma_{ACC0}^2$ is dominant for $q < N < q^{-1/6}$, and $\sigma_{MCC1}^2 > \sigma_{ACC0}^2$ for $q^{-1/6} < N < q^{-1}$. These triangular-shaped areas are marked as ACC0 and MCC1 in Fig. 1. At the line $q = N$, $q < 1$ both the PT result, Eq. (60) and $\sigma_{ACC0}^2$, Eq. (62) are $O(1)$. At the line $q = N^{-1}$, $q < 1$, both the PT result, Eq. (61) and $\sigma_{MCC1}^2$, Eq. (63) are $O(N^{5/3}) > 1$. The merger of these asymptotes confirms that there are no intermediate asymptotes between the PT and CC asymptotes and Eqs. (60 – 63) complemented by the stated validity domains form a complete asymptotic map for $\sigma_{OAM}^2(N, q)$ in the case of non-circular-symmetric apertures, as shown in Fig. 1.

### 2. Circular-symmetric apertures

For a circular-symmetric aperture, the first-order PT term vanishes, and the second-order $\sigma_{PT2}^2$, Eq. (38), becomes the leading asymptotic term for weak turbulence. The two asymptotes $\sigma_{PT2}^2$ for small and large aperture Fresnel numbers are given by Eqs. (39) and (40).

For the Kolmogorov spectrum $\kappa_{1,2} \propto a^{-1}$ in the integrals of Eq. (39) and Eq. (40), for the second-order perturbation asymptotes, it is straightforward to estimate that for $N < 1$

$$\sigma_{PT2}^2 = O\left[\left(C_n^2 k^2 L a^{5/3}\right)^2\right] = O\left(q^{-5/3} N^{5/3}\right), \ N < 1, \quad (65)$$

And for $N > 1$

$$\sigma_{PT2}^2 = O\left[\left(C_n^2 k L^2 a^{-1/3}\right)^2\right] = O\left(q^{-5/3} N^{-1/3}\right), \quad N > 1. \quad (66)$$

As expected, at the boundary, $N \propto 1$ for both Eq. (62) and Eq. (63) have the same order $\sigma_{PT2}^2 \propto O\left(q^{-5/3}\right)$.

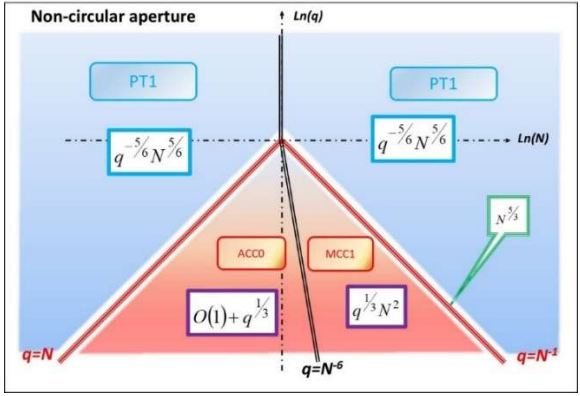

Fig. 1. Asymptotic map for $\sigma_{OAM}^2(N,q)$ in case of non-circular-symmetric apertures.

For circular-symmetric apertures, the two leading terms of the CC expansion are $\sigma_{ACC0}^2$ and $\sigma_{MCC2}^2$. Eq. (62) is still valid for $\sigma_{ACC0}^2$, and here, we can provide a numerical value of this term for a "top hat" aperture as an example

$$\sigma_{ACC0}^2 = \frac{5\pi}{12} 2\pi \int_0^a R^3 dR \left(2\pi \int_0^a R dR\right)^{-2} = \frac{5}{24}. \quad (67)$$

In Eq. (55) $t(z_1, z_2, \zeta) \propto r_C$ as is necessary for the CC series. This implies that in Eq. (55) $\kappa_{1,2} \propto k r_C / L$, $|W| \propto q$, and

$$\sigma_{MCC\,2}^2 = O\left(q^{2/3} N^2\right). \quad (68)$$

Hence the CC estimate of $\sigma_{OAM}^2$ for a circular-symmetric aperture is

$$\sigma_{OAM}^2 = \sigma_{ACC0}^2 + \sigma_{MCC1}^2 = O(1) + O\left(q^{2/3} N^2\right). \quad (69)$$

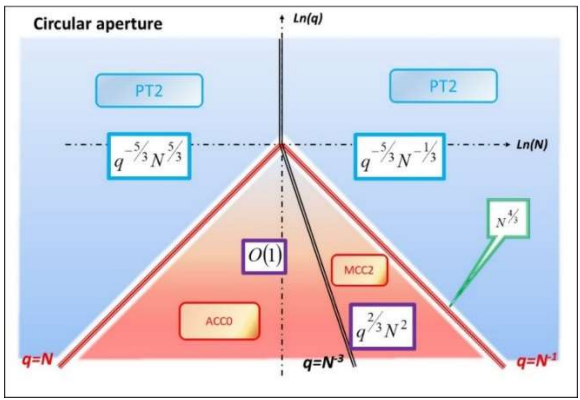

Fig. 2. Asymptotic map for $\sigma_{OAM}^2(N,q)$ in the case of circular-symmetric apertures.

From the derivation of the terms of the CC expansion, we should expect that Eq. (69) is valid in the domain given by Eq. (49), which is shown as an equilateral triangle at the lower part of Fig. 2. In this domain $\sigma_{ACC0}^2$ is dominant for $q < N < q^{-1/3}$, and $\sigma_{MCC2}^2 > \sigma_{ACC0}^2$ for $q^{-1/3} < N < q^{-1}$. These triangular-shaped areas are marked as ACC0 and MCC2 in Fig. 2. At the line $q = N$, $q < 1$ both PT result, Eq. (65) and $\sigma_{ACC0}^2$, Eq. (62) are $O(1)$. At the line $q = N^{-1}$, $q < 1$ both PT result, Eq. (66) and $\sigma_{MCC2}^2$, Eq. (68) are $O\left(N^{5/3}\right) > 1$. This confirms that there are no intermediate asymptotes between the PT and CC asymptotes and Eqs. (65, 66, 62, 68) with specified validity domains form a complete asymptotic map for $\sigma_{OAM}^2(N,q)$ in the case of circular-symmetric apertures, as shown in Fig. 2.

## 4. DISCUSSION

Even without calculating the numerical coefficients in the asymptotic equations, we can make some important qualitative conclusions regarding the dependence of $\sigma_{OAM}^2$ on the path length, turbulence strength, wave length and aperture size.

- Weak and strong fluctuation conditions for the aperture-integrated OAM are not determined by the "Rytov variance," Eq. (27). For small apertures with an aperture size smaller than the Fresnel radius, $a < \sqrt{\lambda L}$ the weak/strong fluctuation boundary is determined by the ratio of the aperture size $a$ to the coherence radius $r_C$, or the ratio $q/N$. For large apertures with an aperture size larger that the Fresnel zone $a > \sqrt{\lambda L}$ the weak/strong fluctuation boundary is determined by the ratio of the aperture size to the so-called "scattering disk," $L/kr_C$, or the product $qN$. These boundaries are the same for both circular-symmetric and non-circular-symmetric apertures.

- The case of very large apertures corresponds to the far right-hand side of the asymptotic maps in Figs. 1 and 2. One should expect some aperture-averaging effect when the local OAM density fluctuations are suppressed by integration over the aperture area that is proportional to Fresnel number $N$. Our results provide $\sigma_{OAM}^2 \propto N^{5/6}$ for non-circular-symmetric apertures, Eq. (61), and $\sigma_{OAM}^2 \propto N^{-1/3}$ for circular-symmetric apertures, Eq. (66). This clearly differs from the textbook "$\sqrt{n}$ law," which at first glance should describe the fluctuation averaging of the OAM density fluctuations by the aperture. If the "$\sqrt{n}$ law" holds, then $\sigma_{OAM}^2 \propto N^{-1}$. Note, however, that the OAM density is not a statistically homogeneous field, since $\langle M^2(r)\rangle \propto r^2$, Eq. (34). Assuming that $M(r)/r$ is homogeneous would result in $\sigma_{OAM}^2 \propto N$, which is still not the case for both aperture classes. It is also clear that the large-aperture asymptotes, Eqs. (61) and (66), depend on the shape of the turbulence spectrum, as is evidenced by the exponents of the $N^{5/6}$ and $N^{-1/3}$ power law dependencies. A similar situation arises for the scintillation averaging [13] where the power flux variance is proportional to the $(\text{area})^{-7/6}$ instead of $(\text{area})^{-1}$. In the latter case, this "super-averaging" [17] effect is related to the energy conservation, however the cause of the OAM fluctuation averaging behavior requires further investigation.

- In the case when the incident wave scintillations at the aperture are weak, $q > 1$, or $\sigma_I^2 < 1$, the OAM variance dependence on the aperture size is notably different for circular-symmetric and asymmetric apertures. For non-circular-symmetric apertures $\sigma_{OAM}^2 \propto a^{5/3}$ for both small, $a < \sqrt{L/k}$ and large, $a > \sqrt{L/k}$ apertures with a transitional range at $a \propto \sqrt{L/k}$. This dependence is presented in a schematic manner in Fig. 3. It is remarkable that $\sigma_{OAM}^2$ can reach very large values, while still described by the perturbation theory.

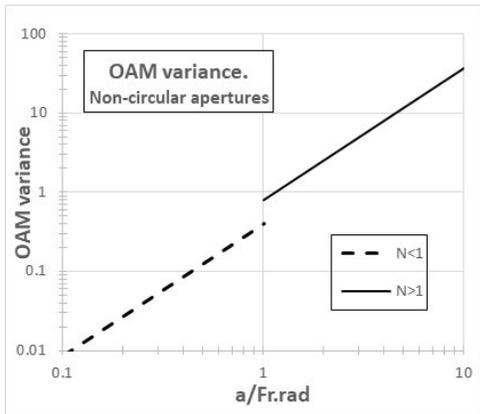

Fig. 3. Dependence of $\sigma_{OAM}^2$ on the normalized aperture size $a/\sqrt{L/k}$ for a non-circular-symmetric aperture in the weak fluctuation case.

For circular-symmetric apertures, $\sigma_{OAM}^2$ grows rapidly with the aperture size as $a^{10/3}$ when the aperture size is smaller than the Fresnel radius, $a < \sqrt{L/k}$, reaches its maximum $\max(\sigma_{OAM}^2) = O\left[\left(\sigma_I^2\right)^2\right]$ at $a \propto \sqrt{L/k}$, then decays as $a^{-2/3}$ for $a > \sqrt{L/k}$. This behavior is illustrated in Fig. 4.

It is notable that for non-circular-symmetric apertures $\sigma_{OAM}^2 \propto \sigma_I^2$, but for circular-symmetric apertures $\sigma_{OAM}^2 \propto \left(\sigma_I^2\right)^2$. In general, OAM fluctuations are smaller for circular-symmetric apertures.

- For small non-circular-symmetric apertures, $a < \sqrt{L/k}$ dependence of the $\sigma_{OAM}^2$ on the turbulence strength measured by the "Rytov variance," $\sigma_I^2$ Eq. (27), is shown in Fig. 5. The aperture Fresnel number for both plots is $N \approx 0.2$. In this case, $\sigma_{OAM}^2$ growth as $\sigma_I^2$ until the coherence radius becomes close to the aperture size, and then saturates at a constant value that depends only on the aperture shape, Eq. (62). In Fig. 5, the saturation value was arbitrary chosen to be unity.

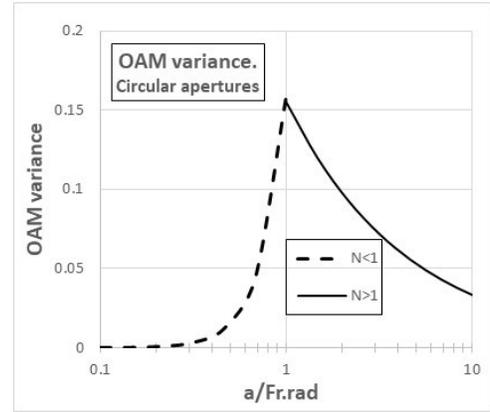

Fig. 4. Dependence of $\sigma_{OAM}^2$ on the normalized aperture size $a/\sqrt{L/k}$ for a circular-symmetric aperture in the weak fluctuation case.

- For small circular-symmetric apertures, $\sigma_{OAM}^2$ growth as $\left(\sigma_I^2\right)^2$, and saturates for $r_C < a$ at the $\sigma_{ACC0}^2$ value, Eq. (62), as shown in Fig. 6. In this figure, we used the accurate $5/24$, Eq. (67), saturation value for a "top hat" circular aperture.

- The dependence of the $\sigma_{OAM}^2$ on $\sigma_I^2$ is shown in Fig. 7 for large non-circular-symmetric apertures, $a > \sqrt{L/k}$. The aperture Fresnel number is $N \approx 16$. In this case $\sigma_{OAM}^2$ grows as $\sigma_I^2$ until it peaks at $\max(\sigma_{OAM}^2) = O(N^{5/3})$, when the scattering disk becomes close to the aperture size, $a \propto L/kr_C$. Thereafter, following the MCC1 asymptote, Eq. (63), $\sigma_{OAM}^2$ decays as $\left(\sigma_I^2\right)^{-2/5}$. Eventually, for $\sigma_I^2 > N^5$, $\sigma_{OAM}^2$ saturates at the constant value $\sigma_{ACC0}^2$. However, this saturation happens at extremely large values of $\sigma_I^2$, that are outside the chart range.

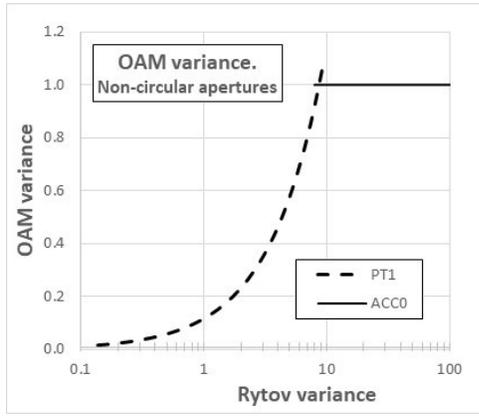

Fig. 5. Dependence of $\sigma^2_{OAM}$ on $\sigma^2_I$ for small non-circular-symmetric apertures, $N \approx 0.2$.

For the large circular-symmetric aperture case shown in Fig. 8, the growth of $\sigma^2_{OAM}$ is much slower, $\propto (\sigma^2_I)^2$, until it peaks at $\max(\sigma^2_{OAM}) = O(N^{4/3})$, when $a \propto L/kr_C$. Thereafter, following the MCC2 asymptote, Eq. (68), $\sigma^2_{OAM}$ decays as $(\sigma^2_I)^{-4/5}$. Eventually, for $\sigma^2_I > N^{5/2}$, $\sigma^2_{OAM}$ saturates at the constant value $\sigma^2_{ACC0}$ shown as the dotted line in Fig. 8.

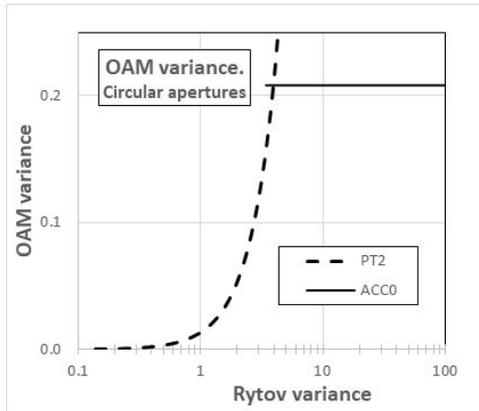

Fig. 6. Dependence of $\sigma^2_{OAM}$ on $\sigma^2_I$ for small circular-symmetric apertures, $N \approx 0.2$.

Our objective here was to provide only the order-of-magnitude estimates of the leading asymptotic terms. In most cases, the numerical coefficients for asymptotes depend on the aperture shape and require numerical integration of multifold integrals. Consequently, the charts shown in Figs. 3–9 are intended only as an illustration of the trends predicted by the asymptotic formulas. In reality, the sharp break points at the intersection of the asymptotic curves are replaced by smooth transitions between the two asymptotic regimes. The specific values of $\sigma^2_{OAM}$ from these charts should not be used to predict the observed OAM fluctuations, with the exception of the strong turbulence saturation value $\sigma^2_{OAM} = 5/24$ for a large "top hat" aperture under very strong turbulence conditions.

## 5. CONCLUSIONS

Based on the general theory of the evolution of the OAM during propagation through turbulence presented in [1], we developed equations for the mean OAM density and OAM density variance for a spherical wave propagating through a random inhomogeneous medium.

The mean OAM density is unaffected by the medium inhomogeneity, and remains at zero for the spherical wave example which was considered. The OAM density variance is related to the fourth-order coherence function, and was calculated only in the first order of the perturbation theory. Its growth occurs as $r^2$ with the distance from the reference point, and is sensitive to the inner scale of turbulence.

The main objective of this work is the estimate of the variance of the aperture-integrated OAM, $\sigma^2_{OAM}$. We used an asymptotic analysis of the fourth-order coherent function in the Feynman path integral representation to explore the dependence of $\sigma^2_{OAM}$ on the turbulence strength, wavenumber, path length and the aperture size for both weak and strong fluctuations.

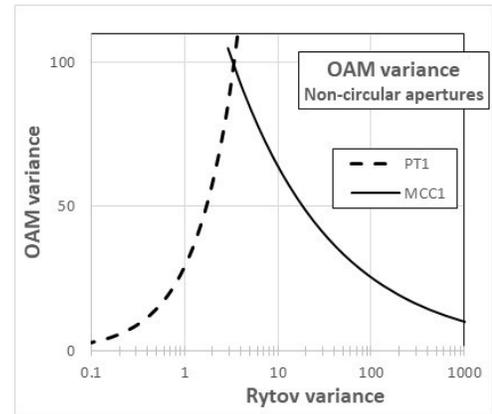

Fig. 7. Dependence of $\sigma^2_{OAM}$ on $\sigma^2_I$ for large non-circular-symmetric apertures, $N \approx 16$.

In the case of Kolmogorov turbulence, $\sigma^2_{OAM}$ depends on two dimensionless parameters which are the Fresnel numbers for the aperture size and the coherence radius. We found that a combination of perturbation theory and the coherence channel expansion is sufficient to develop complete asymptotic maps of $\sigma^2_{OAM}$. These maps are shown in Figures 1 and 2.

An analysis of the asymptotic maps revealed that there is substantial difference between the aperture-integrated intrinsic OAM fluctuations for the circular-symmetric and non-circular symmetric apertures. In particular, the second-orders of the perturbation theory and coherence channels expansion need to be accounted for in the former. This is an indication of its relatively slow growth of OAM fluctuations with the turbulence strength $\sigma^2_{OAM} \propto (C^2_n)^2$, for circular-symmetric apertures.

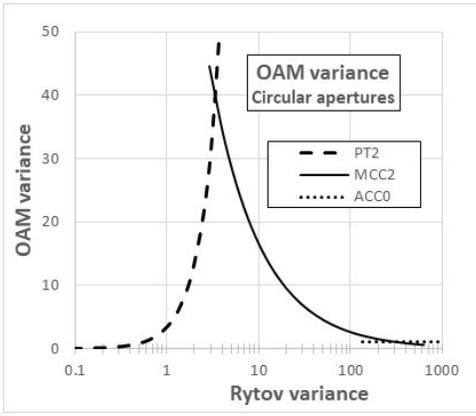

Fig. 8. Dependence of $\sigma^2_{OAM}$ on $\sigma^2_I$ for large circular-symmetric apertures, $N \approx 16$.

We also found that the Rytov variance is not suitable for parameterization of the variance of the aperture-averaged OAM due to the strong dependence on the aperture Fresnel number.

Aperture-averaging of the OAM fluctuations does not obey the standard "square root" law, and the rate of $\sigma^2_{OAM}$ change with the aperture area depends on the turbulence spectrum.

Aperture-averaging of the OAM fluctuations does not obey the standard "square root" law, and the rate of $\sigma^2_{OAM}$ change with the aperture area depends on the turbulence spectrum.

While our results in their present form do not have a predictive capacity for specific $\sigma^2_{OAM}$ values, they identify several distinct propagation regimes that can be observed in experiments and direct propagation modeling.

The aperture-averaged OAM can be measured using a conventional Shack–Hartman wave front sensor. An attempt to compare these theoretical results to the modeling results and field measurements was reported in [8], but more extensive measurements are needed to support the theoretical results presented.

## APPENDIX. PATH INTEGRAL FORMULATION

The solution of the parabolic equation for the fourth-order coherence function, Eq. (23) can be presented in the Feynman path integral form [7, 14]. For a point source located at $(\mathbf{0}, L)$ we have

$$\Gamma_4(\mathbf{r}_1, \boldsymbol{\rho}_1, \mathbf{r}_2, \boldsymbol{\rho}_2, L) = \frac{k^2}{4\pi^2 L^2} \exp\left[\frac{ik}{L}(\mathbf{r}_1 \cdot \boldsymbol{\rho}_1 + \mathbf{r}_2 \cdot \boldsymbol{\rho}_2)\right]$$

$$\times \iint D^2 v_1(\zeta) \iint D^2 v_2(\zeta) \exp\left(ik \int_0^L dz \mathbf{v}_1(z) \cdot \mathbf{v}_2(z)\right)$$

$$\times \delta\left(\int_0^L dz \mathbf{v}_1(z)\right) \delta\left(\int_0^L dz \mathbf{v}_2(z)\right) \quad (A1)$$

$$\times \exp\left[-\frac{\pi k^2}{4} \int_0^L dz \Psi[\mathbf{r}_1(z), \boldsymbol{\rho}_1(z), \mathbf{r}_2(z), \boldsymbol{\rho}_2(z)]\right].$$

Here,

$$\tilde{\mathbf{r}}_1(z) = \mathbf{r}_1\left(1 - \frac{z}{L}\right) + \frac{1}{2}\int_z^L d\zeta \mathbf{v}_1(\zeta),$$

$$\tilde{\mathbf{r}}_2(z) = \mathbf{r}_2\left(1 - \frac{z}{L}\right) - \frac{1}{2}\int_z^L d\zeta \mathbf{v}_1(\zeta),$$

$$\tilde{\boldsymbol{\rho}}_1(z) = \boldsymbol{\rho}_1\left(1 - \frac{z}{L}\right) + \int_z^L d\zeta \mathbf{v}_2(\zeta), \quad (A2)$$

$$\tilde{\boldsymbol{\rho}}_2(z) = \boldsymbol{\rho}_2\left(1 - \frac{z}{L}\right) - \int_z^L d\zeta \mathbf{v}_2(\zeta)$$

are paths (virtual rays) connecting the source point and observation point. The integration in Eq. (A1) is over the space of all continuous paths, but, as discussed in [14] it is presented in the form of an integration over the paths' slopes $\mathbf{v}_{1,2}(\zeta)$.

There are two known asymptotic cases when the path integral in Eq. (A1) can be evaluated analytically [15, 12, 13, 16]. In the weak fluctuation case, the perturbation series for $\Gamma_4$ can be developed by a Taylor expansion of the last exponent in Eq. (A1)

$$\exp\left(-\frac{\pi k^2}{4}\int_0^L dz \Psi\right) \cong 1 - \frac{\pi k^2}{4}\int_0^L dz \Psi + \frac{1}{2}\left(\frac{\pi k^2}{4}\int_0^L dz \Psi\right)^2 + \ldots \quad (A3)$$

In the strong fluctuation case, the Coherence Channel (CC) expansion of the last exponent in Eq. (A1) is used [15, 12]. Namely, $\Gamma_4$ is presented as a sum of the contribution of the Main Coherence Channel (ACC), where the last exponent in Eq. (A1) is approximated as

$$\exp\left(-\frac{\pi k^2}{4}\int_0^L dz \Psi\right)_{MCC} \cong \exp\left(-\frac{\pi k^2}{4}\int_0^L d\zeta [H(\tilde{\boldsymbol{\rho}}_1) + H(\tilde{\boldsymbol{\rho}}_2)]\right)$$

$$\left\{1 + \frac{\pi k^2}{4}\int_0^L dz\left[H\left(\tilde{\mathbf{r}}_1 - \tilde{\mathbf{r}}_2 + \frac{\tilde{\boldsymbol{\rho}}_1 - \tilde{\boldsymbol{\rho}}_2}{2}\right) + H\left(\tilde{\mathbf{r}}_1 - \tilde{\mathbf{r}}_2 - \frac{\tilde{\boldsymbol{\rho}}_1 - \tilde{\boldsymbol{\rho}}_2}{2}\right)\right.\right. \quad (A4)$$

$$\left.\left. - H\left(\tilde{\mathbf{r}}_1 - \tilde{\mathbf{r}}_2 + \frac{\tilde{\boldsymbol{\rho}}_1 + \tilde{\boldsymbol{\rho}}_2}{2}\right) - H\left(\tilde{\mathbf{r}}_1 - \tilde{\mathbf{r}}_2 - \frac{\tilde{\boldsymbol{\rho}}_1 + \tilde{\boldsymbol{\rho}}_2}{2}\right)\right] + \ldots\right\},$$

and Additional Coherence Channel (ACC), where

$$\exp\left(-\frac{\pi k^2}{4}\int_0^L dz \Psi\right)_{ACC}$$

$$\cong \exp\left\{-\frac{\pi k^2}{4}\int_0^L d\zeta\left[H\left(\tilde{\mathbf{r}}_1 - \tilde{\mathbf{r}}_2 + \frac{\tilde{\boldsymbol{\rho}}_1 + \tilde{\boldsymbol{\rho}}_2}{2}\right) + H\left(\tilde{\mathbf{r}}_1 - \tilde{\mathbf{r}}_2 - \frac{\tilde{\boldsymbol{\rho}}_1 + \tilde{\boldsymbol{\rho}}_2}{2}\right)\right]\right\} \quad (A5)$$

$$\left\{1 + \frac{\pi k^2}{4}\int_0^L dz\left[H\left(\tilde{\mathbf{r}}_1 - \tilde{\mathbf{r}}_2 - \frac{\tilde{\boldsymbol{\rho}}_1 - \tilde{\boldsymbol{\rho}}_2}{2}\right)\right.\right.$$

$$\left.\left. + H\left(\tilde{\mathbf{r}}_1 - \tilde{\mathbf{r}}_2 + \frac{\tilde{\boldsymbol{\rho}}_1 - \tilde{\boldsymbol{\rho}}_2}{2}\right) - H(\tilde{\boldsymbol{\rho}}_1) - H(\tilde{\boldsymbol{\rho}}_2)\right] + \ldots\right\}.$$

It is crucial that, in order for the CC expansion to be a valid approximation in any specific problem related to $\Gamma_4$, the exponential terms in Eqs. (A4) and (A5) have to effectively restrict the integration domains in the applicable integrals. This, somewhat vague for the general case, constraint becomes more specific in calculations of $\sigma_{OAM}^2$ in Subsection 3.2.

Both the perturbation and the CC expansions allow term-by-term calculation of the applicable path integrals and result in closed-form formulations for individual series terms. We refer to the technical details in [12] and present only the necessary final results here.

The zero-order MCC and ACC terms are

$$\Gamma_4^{(MCC\,0)}(\mathbf{r}_1, \boldsymbol{\rho}_1, \mathbf{r}_2, \boldsymbol{\rho}_2) = \Gamma(\mathbf{r}_1, \boldsymbol{\rho}_1, L)\Gamma(\mathbf{r}_2, \boldsymbol{\rho}_2, L), \quad (A6)$$

$$\Gamma_4^{(ACC\,0)}(\mathbf{r}_1, \boldsymbol{\rho}_1, \mathbf{r}_2, \boldsymbol{\rho}_2) = \Gamma_4^{(0)}(\mathbf{r}_1, \boldsymbol{\rho}_1, \mathbf{r}_2, \boldsymbol{\rho}_2)$$

$$\times\left\{-\frac{\pi k^2}{4}\int_0^L dz\left[H\left(\left(\mathbf{r}_1 - \mathbf{r}_2 + \frac{\boldsymbol{\rho}_1 + \boldsymbol{\rho}_2}{2}\right)\left(1 - \frac{z}{L}\right)\right)\right.\right. \quad (A7)$$

$$\left.\left. + H\left(\left(\mathbf{r}_1 - \mathbf{r}_2 - \frac{\boldsymbol{\rho}_1 + \boldsymbol{\rho}_2}{2}\right)\left(1 - \frac{z}{L}\right)\right)\right]\right\}.$$

The higher-order term become increasingly unwieldly, and we show only the first and second-order MCC terms that are necessary for the OAM variance calculations.

$$\Gamma_4^{(MCC\,1)}(\mathbf{r}_1, \boldsymbol{\rho}_1, \mathbf{r}_2, \boldsymbol{\rho}_2) = 4\pi k^2 \Gamma_4^{(0)}(\mathbf{r}_1, \boldsymbol{\rho}_1, \mathbf{r}_2, \boldsymbol{\rho}_2)\int_0^L dz \iint d^2\kappa \Phi_n(\boldsymbol{\kappa})$$

$$\times \exp\left\{i\boldsymbol{\kappa}\cdot(\mathbf{r}_1 - \mathbf{r}_2)\left(1 - \frac{z}{L}\right) - \frac{\pi k^2}{4}\int_0^L d\zeta(H[\mathbf{p}_1(z,\zeta)] + H[\mathbf{p}_2(z,\zeta)])\right\} \quad (A8)$$

$$\times\left\{\cos\left[\frac{\boldsymbol{\kappa}}{2}\cdot(\mathbf{p}_1(z,z) + \mathbf{p}_2(z,z))\right] - \cos\left[\frac{\boldsymbol{\kappa}}{2}\cdot(\mathbf{p}_1(z,z) - \mathbf{p}_2(z,z))\right]\right\},$$

where single-kink paths are

$$\mathbf{p}_1(z,\zeta) = \boldsymbol{\rho}_1\left(1 - \frac{\zeta}{L}\right) - \frac{\boldsymbol{\kappa}}{k}\min(z,\zeta)\zeta\left[1 - \frac{\max(z,\zeta)}{L}\right],$$

$$\mathbf{p}_2(z,\zeta) = \boldsymbol{\rho}_2\left(1 - \frac{\zeta}{L}\right) + \frac{\boldsymbol{\kappa}}{k}\min(z,\zeta)\left[1 - \frac{\max(z,\zeta)}{L}\right]. \quad (A9)$$

And

$$\Gamma_4^{(MCC\,2)}(\mathbf{r}_1, \boldsymbol{\rho}_1, \mathbf{r}_2, \boldsymbol{\rho}_2) = 32\pi^2 k^4 \Gamma_4^{(0)}(\mathbf{r}_1, \boldsymbol{\rho}_1, \mathbf{r}_2, \boldsymbol{\rho}_2)$$

$$\int_0^L dz_1 \int_0^L dz_2 \iint d^2\kappa_1 \Phi_n(\boldsymbol{\kappa}_1) \iint d^2\kappa_2 \Phi_n(\boldsymbol{\kappa}_2)$$

$$\times \exp\left\{i\left[\boldsymbol{\kappa}_1\left(1 - \frac{z_1}{L}\right) + \boldsymbol{\kappa}_2\left(1 - \frac{z_2}{L}\right)\right]\cdot(\mathbf{r}_1 - \mathbf{r}_2)\right\}$$

$$\times \exp\left\{-\frac{\pi k^2}{4}\int_0^L d\zeta(H[\mathbf{q}_1(z_1,z_2,\zeta)] + H[\mathbf{q}_2(z_1,z_2,\zeta)])\right\} \quad (A10)$$

$$\times \cos\left[\frac{1}{2}\boldsymbol{\kappa}_1\cdot\mathbf{q}_1(z_1,z_2,z_1)\right]\cos\left[\frac{1}{2}\boldsymbol{\kappa}_1\cdot\mathbf{q}_2(z_1,z_2,z_1)\right]$$

$$\times \cos\left[\frac{1}{2}\boldsymbol{\kappa}_2\cdot\mathbf{q}_1(z_1,z_2,z_2)\right]\cos\left[\frac{1}{2}\boldsymbol{\kappa}_2\cdot\mathbf{q}_2(z_1,z_2,z_2)\right],$$

where two-kinks paths are

$$\mathbf{q}_1(z_1,z_2,\zeta) = \boldsymbol{\rho}_1\left(1 - \frac{\zeta}{L}\right) - \frac{\boldsymbol{\kappa}_1}{k}\min(z_1,\zeta)\zeta\left[1 - \frac{\max(z_1,\zeta)}{L}\right]$$

$$- \frac{\boldsymbol{\kappa}_2}{k}\min(z_2,\zeta)\zeta\left[1 - \frac{\max(z_2,\zeta)}{L}\right], \quad (A11)$$

$$\mathbf{q}_2(z_1,z_2,\zeta) = \boldsymbol{\rho}_2\left(1 - \frac{\zeta}{L}\right) + \frac{\boldsymbol{\kappa}_1}{k}\min(z_1,\zeta)\zeta\left[1 - \frac{\max(z_1,\zeta)}{L}\right]$$

$$+ \frac{\boldsymbol{\kappa}_2}{k}\min(z_2,\zeta)\zeta\left[1 - \frac{\max(z_2,\zeta)}{L}\right].$$

It is important to mention that Eqs. (A8) and (A10) are not the final forms of the corresponding series terms. As was mentioned earlier, the notion of the CC series requires that the paths $\mathbf{p}_{1,2}(z,\zeta)$ and $\mathbf{q}_{1,2}(z_1,z_2,\zeta)$ are restrained by CC having a width on the order of the coherence radius $r_C$, Eq. (27).